\documentclass[lettersize,journal]{IEEEtran}
\usepackage{amsmath,amsfonts}
\usepackage{algorithmic}
\usepackage{algorithm}
\usepackage{array}
\usepackage[caption=false,font=normalsize,labelfont=sf,textfont=sf]{subfig}
\usepackage{textcomp}
\usepackage{hyperref}
\usepackage{stfloats}
\usepackage{url}
\usepackage{verbatim}
\usepackage{graphicx}
\usepackage{cite}
\usepackage{romannum}
\usepackage{multirow}
\usepackage{booktabs}
\begin{document}
\thispagestyle{empty}
\onecolumn
\null
\vspace*{\fill}
{\large\noindent This work has been accepted for publication in the IEEE Internet of Things Journal \\(\href{https://doi.org/10.1109/JIOT.2025.3567543}{doi.org/10.1109/JIOT.2025.3567543}).

\vspace{\baselineskip}

\noindent\textbf{Copyright:} © 2025 IEEE.  Personal use of this material is permitted.  Permission from IEEE must be obtained for all other uses, in any current or future media, including reprinting/republishing this material for advertising or promotional purposes, creating new collective works, for resale or redistribution to servers or lists, or reuse of any copyrighted component of this work in other works.}
\vspace*{\fill}
\vspace*{\fill}
\vspace*{\fill}

\clearpage
\twocolumn
\setcounter{page}{1}
\title{Noise-Aware Ensemble Learning for Efficient Radar Modulation Recognition
}

\author{Do-Hyun~Park,
        Min-Wook~Jeon,
        Jinwoo Jeong,
        Isaac Sim,
        Sangbom Yun,
        Junghyun Seo,
        and~Hyoung-Nam~Kim,~\IEEEmembership{Member,~IEEE,}
\thanks{Do-Hyun Park, Min-Wook Jeon, and Hyoung-Nam Kim are with the School of Electrical and Electronics Engineering, Pusan National University, Busan 46241, Republic of Korea.}
\thanks{Jinwoo Jeong, Isaac Sim, Sangbom Yun, and Junghyun Seo are with the Cyber Electronic Warfare Research Institute, LIG Nex1, Seongnam 13488, Republic of Korea.}}

\markboth{Journal of \LaTeX\ Class Files}%
{Shell \MakeLowercase{\textit{et al.}}: Bare Demo of IEEEtran.cls for IEEE Journals}


\maketitle
\begin{abstract}
Electronic warfare support (ES) systems intercept adversary radar signals and estimate various types of signal information, including modulation schemes. The accurate and rapid identification of modulation schemes under conditions of very low signal power remains a significant challenge for ES systems. This paper proposes a recognition model based on a noise-aware ensemble learning (NAEL) framework to efficiently recognize radar modulation schemes in noisy environments. The NAEL framework evaluates the influence of noise on recognition and adaptively selects an appropriate neural network structure, offering significant advantages in terms of computational efficiency and recognition performance. We present the analysis results of the recognition performance of the proposed model based on experimental data. Our recognition model demonstrates superior recognition accuracy with low computational complexity compared to conventional classification models.
\end{abstract}

\begin{IEEEkeywords}
Electronic warfare support, modulation recognition, low probability of intercept, ensemble learning, convolutional neural network
\end{IEEEkeywords}

\section{Introduction}
\begin{figure}[t]
\centering
\includegraphics[width= 0.48 \textwidth]{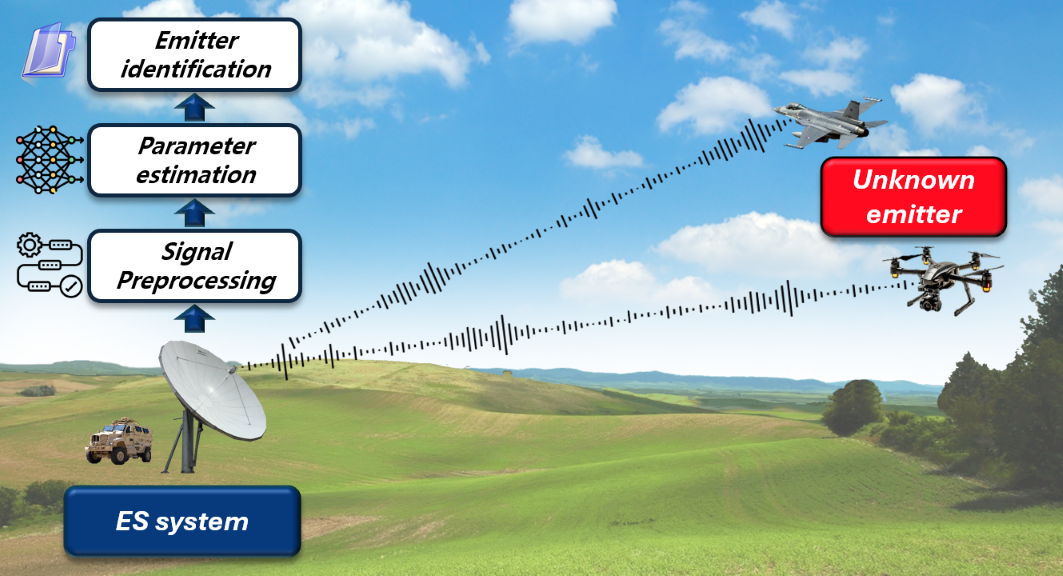}
\caption{Processes of the ES system for emitter identification.}
\label{fig_introduction}
\end{figure}

\IEEEPARstart{M}{odulation} recognition is a key task for identifying the characteristics of unknown received signals and is widely employed in both civilian and military fields. In civilian applications, modulation recognition techniques are primarily aimed at communication signals and are extensively used in cognitive radio and spectrum surveillance\cite{AMCSurvey, AMCSurvey2, AMCSurvey3}. In military applications, modulation recognition is used in electronic warfare support (ES) systems. ES systems estimate the parameter information of unknown radar signals, such as pulse width, time of arrival, center frequency, and modulation scheme. The acquired signal information can be used to identify the type of unknown emitter by comparison with a database stored in the emitter library of the ES systems. The process from signal reception to emitter identification in ES systems is illustrated in Fig. \ref{fig_introduction}. By utilizing this emitter identification information, allied forces can engage in actions such as radio jamming or deception\cite{ELINTBook, PaceBook}.

Recently, many radar emitters utilize low-probability-of-intercept (LPI) radar, having properties that make detection by an opponent's intercept receiver less likely. Therefore, the intercept receivers of ES systems are required to recognize the LPI radar modulation. Among the various features of LPI radars, the most prominent is their low signal-transmission power\cite{PaceBook}. ES systems receive signals with very low signal-to-noise ratios (SNRs) because of their low transmission output. Consequently, extensive research has focused on the robust recognition of modulation schemes in low SNR environments.

In general, modulation recognition techniques involve two processes: signal representation and classification. Various techniques for signal representation include features such as power spectral density\cite{FeaturePSD} and statistical feature extraction using the instantaneous phase and frequency\cite{FeatureIPF}. In addition, time-frequency images (TFIs), such as the short-time Fourier transform\cite{FeatureSTFT}, Wigner-Ville distribution (WVD)\cite{PaceBook,FeaturePSD,CNNWVD}, and Choi-Williams distribution (CWD)\cite{PaceBook,FeaturePSD,FeatureCWD,CNNCWD,CNNCWD1,CNNCWD2,CNNCWD3,EuRAD} are often used. The classification process uses hierarchical decision trees \cite{FeatureIPF,FeatureSTFT,ClassfDT}, support vector machines \cite{ClassfSVM1,ClassfSVM2}, and a multilayer perceptron (MLP)\cite{PaceBook,FeaturePSD,CNNWVD,FeatureCWD,CNNCWD,CNNCWD2,CNNCWD3,EuRAD} to classify the modulation scheme. Recently, convolutional neural network (CNN) models that use convolution layers to extract spatial features from TFIs and employ an MLP (or fully connected layers) for modulation scheme classification have shown improved recognition performance compared with conventional methods, especially in environments with low SNR\cite{CNNWVD,CNNCWD,CNNCWD3}. 

\begin{figure}[t]
\centering
\includegraphics[width= 0.48 \textwidth]{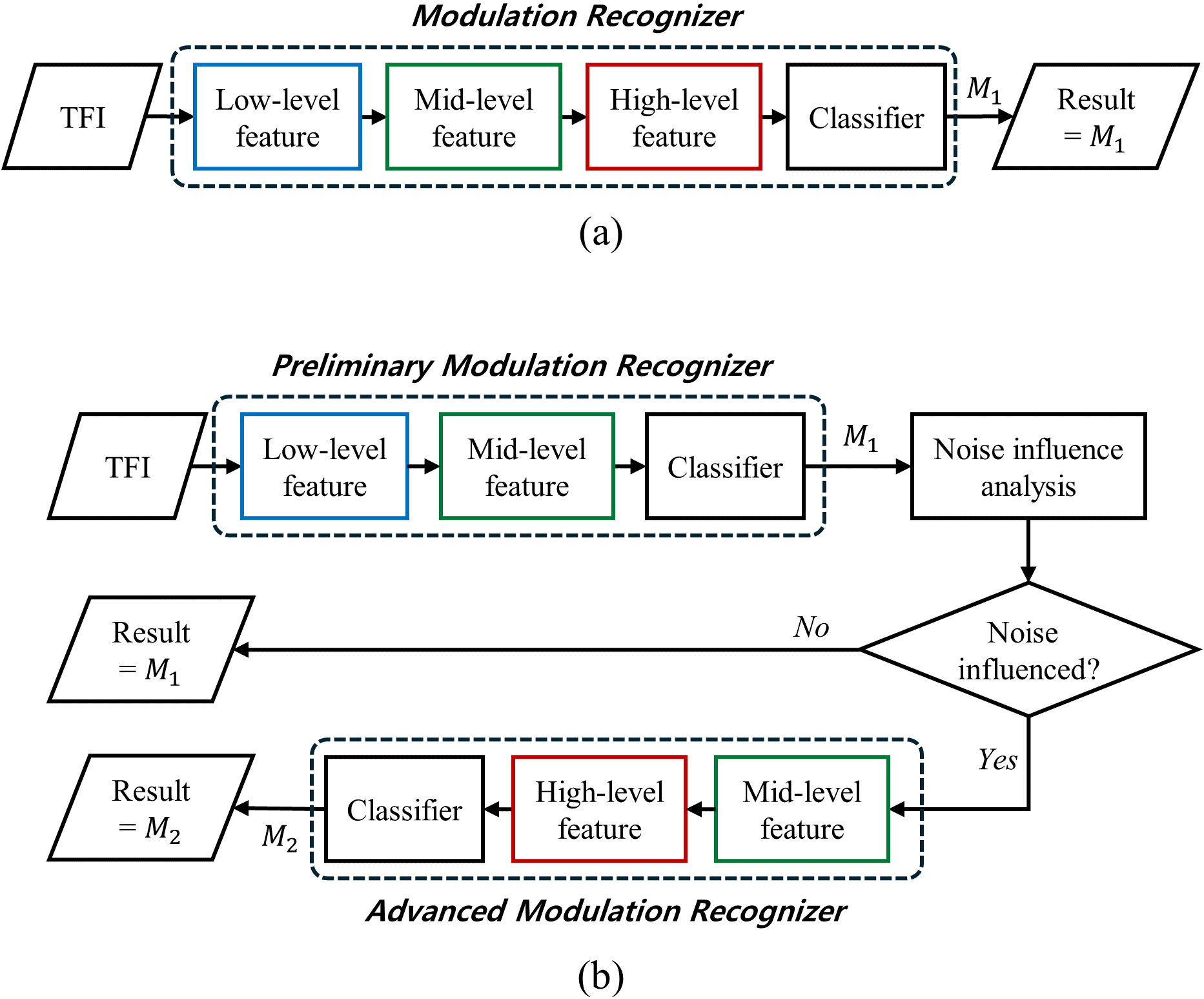}
\caption{Flow charts of modulation recognition models based on CNN: (a) conventional CNN architecture for modulation recognition, (b) proposed CNN architecture for modulation recognition.}
\label{fig_flowchart}
\end{figure}

The conventional CNN-based modulation recognition technique follows the algorithm flowchart shown in Fig. \ref{fig_flowchart}(a). First, a TFI was generated from the received signals, and the spatial features were extracted using convolution layers. During the feature extraction process, the depth of the features extracted by each convolution layer varied. In the initial convolution layers, basic low-level features such as corners and edges were extracted from the input image. In the later stages, these basic features led to the extraction of more complex and abstract mid- and high-level features based on cumulative knowledge from earlier stages\cite{Visualizing}. Finally, a classifier comprising fully connected layers used the extracted features to determine the modulation scheme. 

Because ES systems must immediately estimate the information of the received signals, the use of lightweight CNN models is imperative. Considerable research has been conducted on CNN models that efficiently extract modulation features with low computational complexity while maintaining high recognition accuracy. CNNs that use depth-wise convolution are acknowledged to be more efficient than conventional CNNs, and such models have demonstrated the ability to efficiently classify unknown signal modulation schemes\cite{MobileNetV1,MobileNetV2,MobileNet_AMC1,MobileNet_AMC2}. The neural architecture search (NAS) technique has drawn attention for its ability to automatically design optimized neural network structures, thereby maximizing efficiency and performance in processing complex and varied data\cite{NasNet,MNasNet}. The application of the NAS technique in the field of radar modulation recognition has demonstrated its potential for designing efficient recognition models\cite{MNasNet_AMC1,MNasNet_AMC2}. In addition, a channel-shuffling technique designed to enhance the learning capabilities and parameter efficiency of neural networks has been proposed, and classification models utilizing this technique have shown excellent performance in modulation recognition\cite{ShuffleNetV1,ShuffleNetV2,ShuffleNet_AMC}.

More recently, several advanced deep learning models have been introduced to enhance radar modulation recognition performance further. LPI-Net\cite{CNNCWD} employs a time-frequency analysis-based lightweight feature extraction technique within a deep CNN framework to improve the recognition accuracy of LPI radar waveforms. LWCNN\cite{LWCNN} focuses on robustness against various types of noise, leveraging squeeze-and-excitation networks to achieve improved classification performance under noisy conditions. Additionally, the vision transformer-based recognition model\cite{VIT} incorporates transformer architectures instead of conventional CNNs, allowing for better learning of long-range dependencies in modulation characteristics. 

Although existing CNN-based models exhibit efficiency in radar modulation recognition, a significant issue with these models is their substantial decrease in performance in low SNR environments. For instance, MobileNet, which utilizes depth-wise convolutions, and ShuffleNet, which employs channel shuffling, both adopt lightweight architectures that prioritize computational efficiency; however, they lack the flexibility to adapt the depth of their feature extraction when noise levels increase. This results in a significant drop in recognition accuracy under low-SNR conditions, as their shallow feature representations struggle to capture crucial modulation characteristics. This issue can be addressed by increasing the computational complexity of the recognition model.
However, recognition models applied to ES systems need to estimate signal parameters instantaneously. In high SNR environments, recognition models that enhance feature extraction capabilities with a high computational load can result in unnecessary power consumption and excessively long inference times. Consequently, existing models that use fixed neural network architectures generate unnecessary computations in low-noise environments or exhibit insufficient recognition capabilities in high-noise environments.

This study proposes a novel neural network architecture to address these issues, as depicted in Fig. \ref{fig_flowchart}(b). The proposed architecture exploits a noise-aware ensemble learning (NAEL) technique that evaluates the influence of noise on recognition and adaptively adjusts the network structure accordingly. The proposed model initially utilizes a lightweight preliminary recognizer to perform modulation recognition. When the recognition outcomes of the preliminary recognizer are identified as affected by noise interference, an advanced recognizer capable of extracting deeper features is employed to determine the final modulation recognition outcomes. By employing a conditional decision-making model with this noise-aware approach, unnecessary computation in high SNR environments can be reduced, while recognition accuracy in low SNR environments is enhanced. Compared to existing models, which maintain a fixed lightweight architecture, the proposed model adapts its feature extraction depth dynamically, ensuring optimal trade-offs between computational efficiency and recognition accuracy. This adaptability makes it particularly effective for real-time ES applications, where both performance and efficiency are critical.

The primary objective of this study is to advance ES systems by introducing an efficient neural network architecture that utilizes a novel method to evaluate the influence of noise on recognition decisions. This study demonstrates that integrating a noise-aware method can substantially enhance recognition accuracy compared to existing models and address the challenges posed by high power consumption in deep-learning models. The major contributions of this study are summarized as follows:
\begin{itemize}
\setlength{\itemsep}{1pt}
\setlength{\parskip}{0pt}
\setlength{\parsep}{0pt}
\item {We introduce a recognition model based on spatial feature extraction (FE) blocks that achieve high modulation recognition performance with low computational load. Our model also incorporates a data-reuse strategy to further enhance efficiency.}
\item{We propose a novel methodology for evaluating the impact of noise on modulation recognition. Furthermore, we propose the NAEL framework, which adaptively selects the optimal neural network structure based on the noise influence on recognition.}
\item{Through performance analysis using real measured data, we demonstrate that our proposed model outperforms and is more efficient than existing CNN models.}
\end{itemize}

The remainder of this paper is organized as follows: In Section \Romannum{2}, we introduce the intercepted radar signal model and the time-frequency analysis method. Section \Romannum{3} outlines the proposed modulation recognition model and a method for assessing the significance of noise in recognition results using a noise-aware network. The results of analyzing the performance and computational loads of the proposed model compared to existing models are provided in Section \Romannum{4}, and Section \Romannum{5} presents our conclusions. 

\begin{table}[t]
\centering
\caption{Modulation functions for 12 modulation schemes}
\includegraphics[width= 0.48 \textwidth]{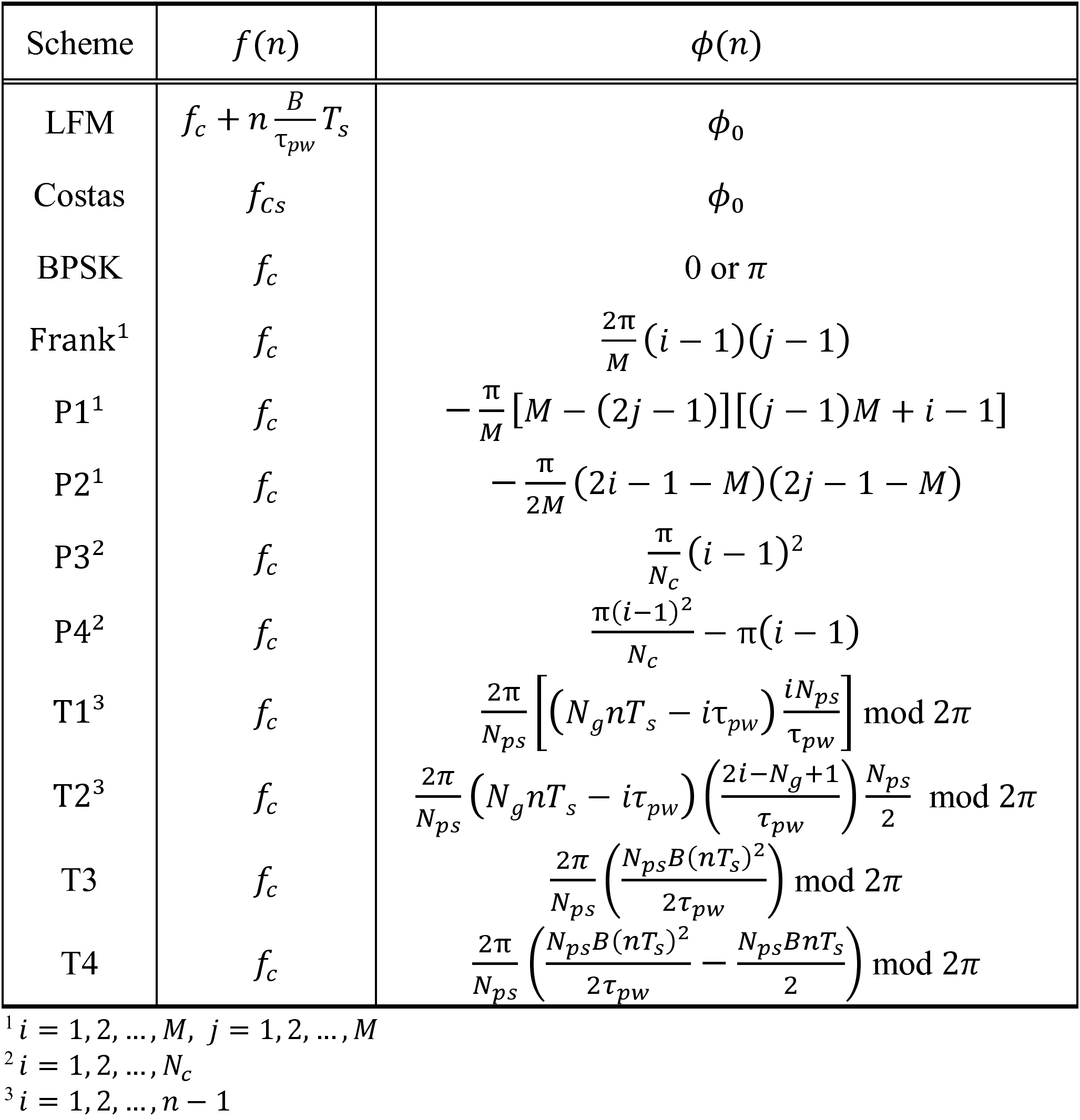}
\label{table_modulation}
\end{table}

\section{Signal Model and Preprocessing}
\subsection{Intercepted Radar Signal}
The discrete-time complex radar signal \(y(n)\) intercepted by the ES receiver can be expressed as
\begin{equation}
y(n)= x(n) + w(n) = A \exp (j 2 \pi f(n) t + \phi(n)) + w(n)
\label{eq_signalmodel}
\end{equation}
where \(n\) is the discrete-time index, \(x(n)\) represents the complex envelopes of the transmitted radar signals, \(w(n)\) represents the complex additive white Gaussian noise, \(A\) is the amplitude, and \(f(n)\) and \(\phi(n)\) are the frequency and phase modulation functions, respectively.

Various modulation techniques can be employed for the radar waveform design. We considered two frequency modulations (FMs) and ten phase modulations (PMs), for a total of 12 modulation schemes for recognition. Specifically, linear FM (LFM) and Costas codes were considered for the FM, and binary phase-shift keying (BPSK), polyphase codes (Frank, P1, P2, P3, and P4), and polytime codes (T1, T2, T3, and T4) were considered for the PM. For the BPSK signal included in the PM, the Barker code was used for the binary code of BPSK.

The phase \(\phi(n)\) in (\ref{eq_signalmodel}) is kept constant with respect to the initial phase \(\phi_{0}\) for FM radar signals, whereas the frequency \(f(n)\) is kept constant at the center frequency \(f_{c}\) for PM radar signals. Table \ref{table_modulation} describes \(f(n)\) and \(\phi(n)\) for the 12 modulation schemes considered in this study, while Table \ref{table_notation} presents the definitions of the parameters used in Table \ref{table_modulation}.

\begin{table}[t]
\centering
\caption{Notation and definitions for modulation functions}
\includegraphics[width= 0.48 \textwidth]{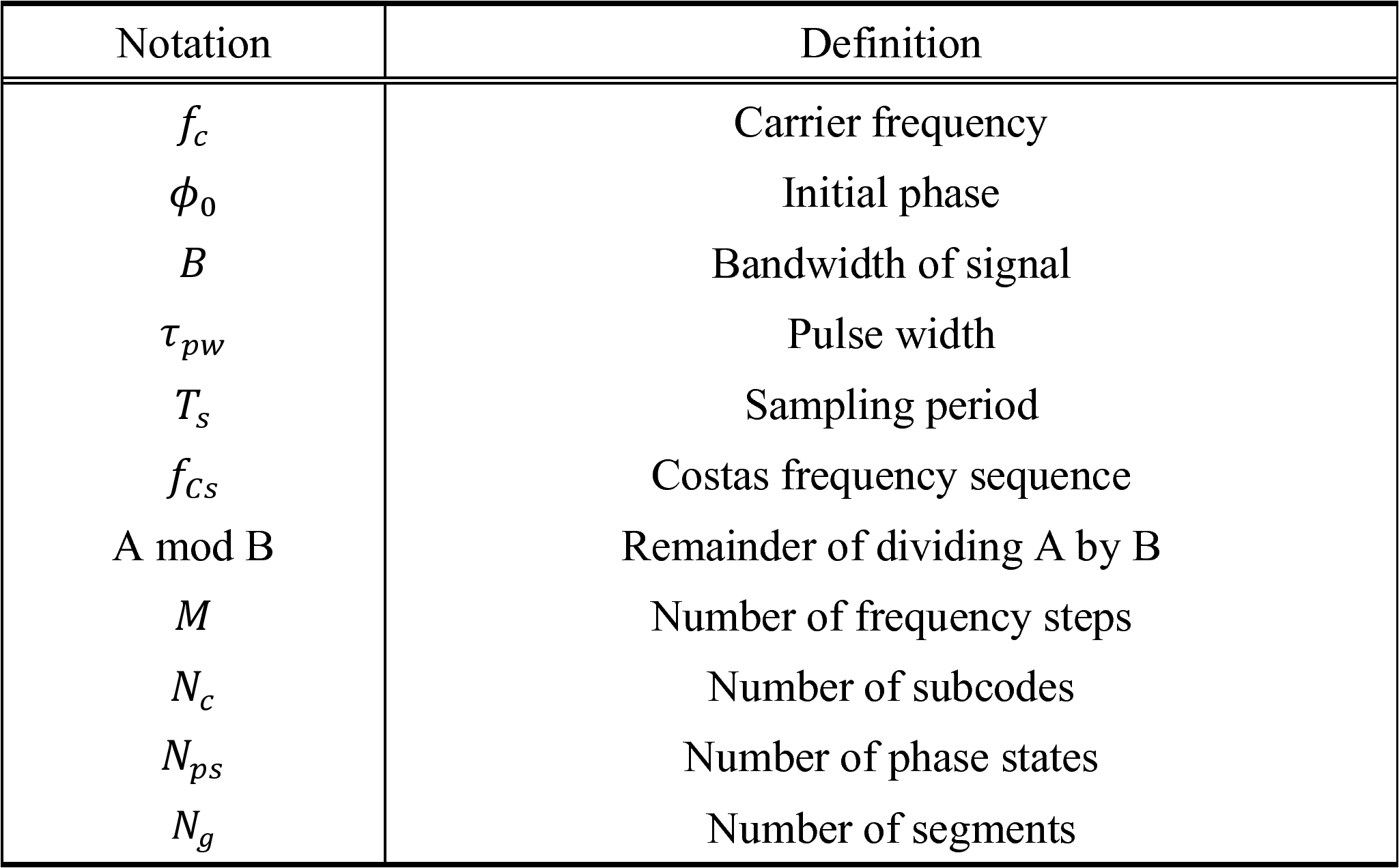}
\label{table_notation}
\end{table}

\subsection{Time-frequency Analysis}
Rather than using the intercepted signal \(y(n)\) directly for modulation recognition, \(y(n)\) is preprocessed before being fed into a neural network to improve recognition performance. Among various signal representation methods, transforming the intercepted signal into a time-frequency image can significantly enhance recognition performance by effectively capturing radar modulation characteristics. As described in Section \Romannum{1}, various time-frequency analysis methods can be employed. This study utilizes the Choi-Williams distribution (CWD), renowned for its exceptional time-frequency analysis performance using an exponential kernel to minimize cross-term components. The continuous CWD of the input signal \(y(t)\) is given by \cite{ChoiWilliams}
\begin{equation}
\begin{split}
\textup{CWD}( & t,f)=\int^{\infty}_{-\infty}\exp(-j\omega\tau) \int^{\infty}_{-\infty} \sqrt{\frac{\sigma}{4 \pi \tau^{2}}} \\
& \exp \left( -\frac{ \left ( \mu - t \right )^{2}}{4 \tau^{2}/\sigma} \right) y\left(\mu+\frac{\tau}{2} \right ) y^{*}\left(\mu-\frac{\tau}{2} \right ) d\mu d\tau
\end{split}
\label{CWD}
\end{equation}
where \(t\) and \(f\) are the time and frequency variables, respectively, \(\omega=2 \pi f\) is the angular frequency, \(\sigma \left( \sigma > 0\right)\) is a scaling factor determining the trade-off between cross-term suppression and frequency resolution. In (\ref{CWD}), the parameter \(\mu\) represents the temporal reference point at which the signal characteristics are analyzed, while \(\tau\) denotes the time lag used to evaluate the local correlation properties of the signal around \(\mu\).

\section{Proposed Modulation Recognition Model}
\begin{figure}[t]
\centering
\includegraphics[width= 0.48 \textwidth]{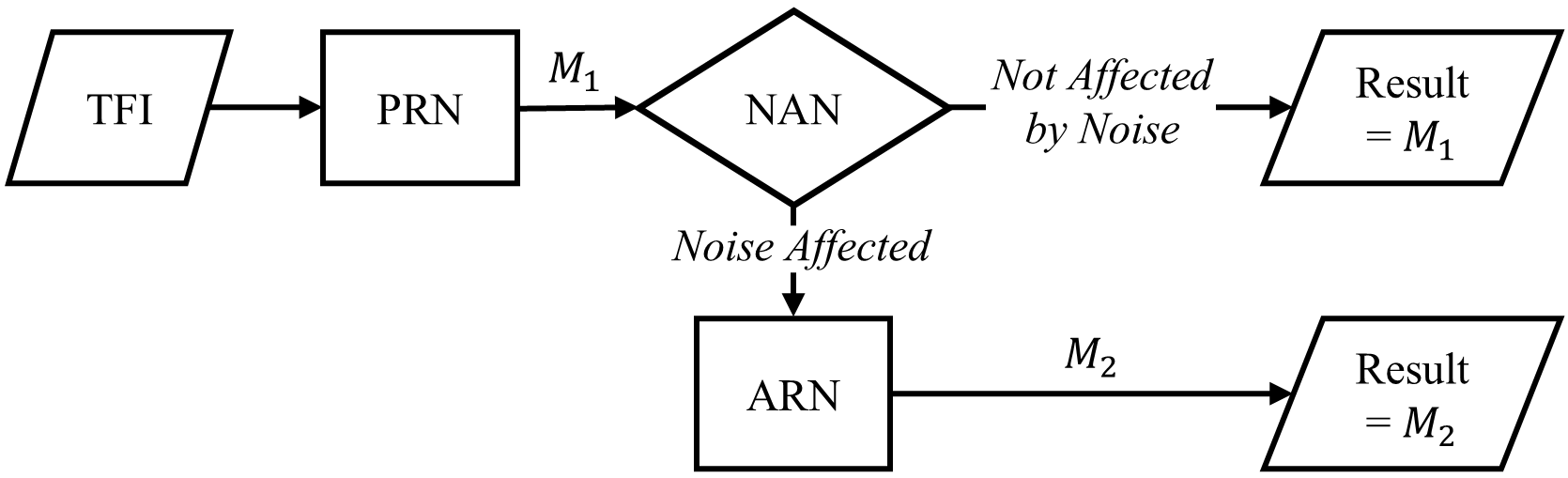}
\caption{A flowchart illustrating the interactions between PRN, NAN, and ARN in the NAEL framework.}
\label{NAELflowchart}
\end{figure}

\begin{figure*}[t]
\centering
\includegraphics[width=\textwidth]{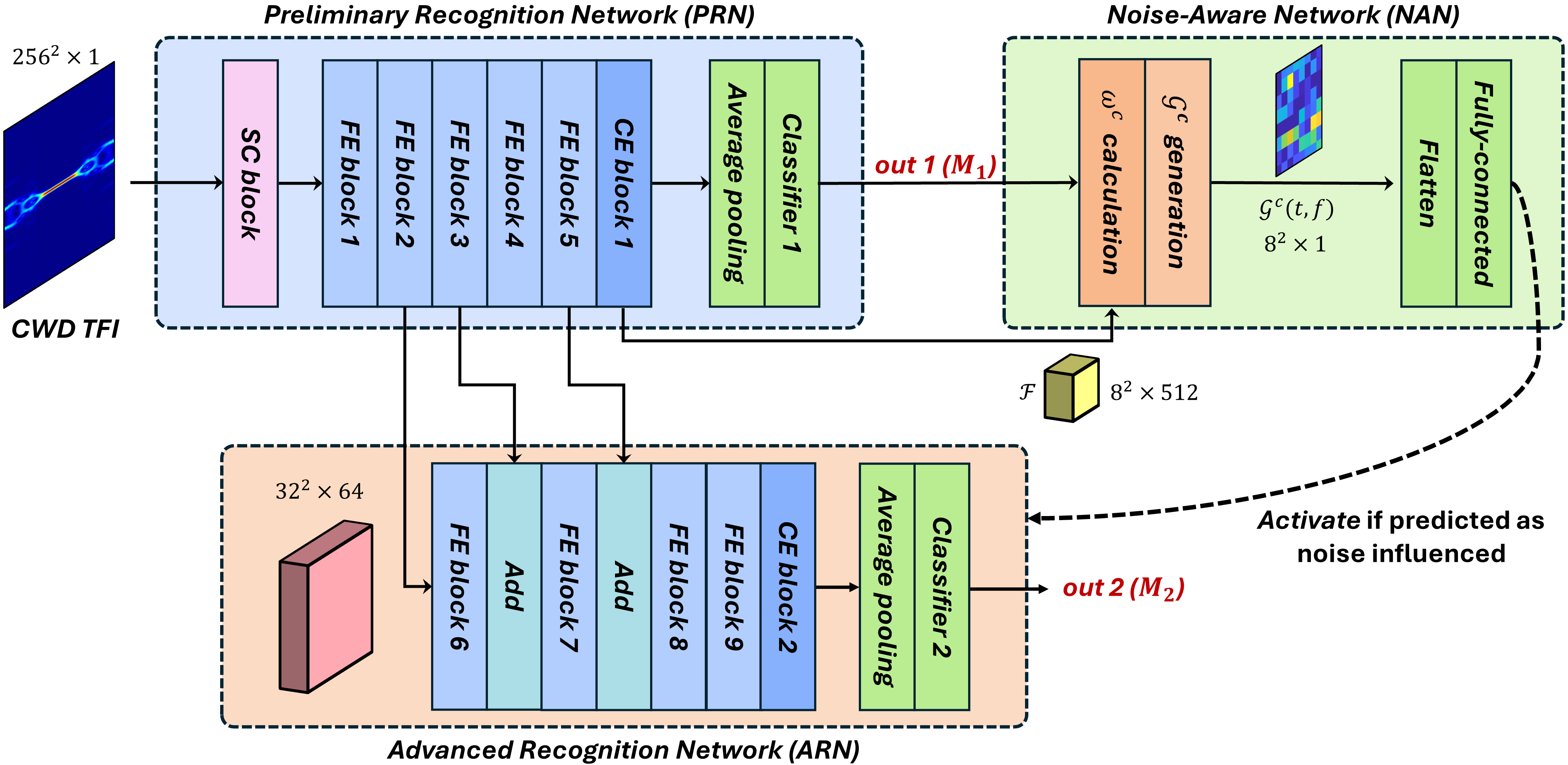}
\caption{Overall architecture of the NAEL framework. Our proposed recognition model uses SC, FE, and CE blocks to produce feature maps and compute importance weights \(w^{c}\) culminating in the creation of a gradient map \(\mathcal{G}^{c}\).}
\label{fig_overall}
\end{figure*}

This section describes the proposed modulation recognition model exploiting the noise-aware ensemble learning (NAEL) framework. This framework consists of three neural networks: the preliminary recognition network (PRN), noise-aware network (NAN), and advanced recognition network (ARN). Their respective roles are summarized in Table \ref{table_acronym}. The interaction and decision-making flow among the PRN, NAN, and ARN are illustrated clearly in Fig. \ref{NAELflowchart}. Furthermore, Fig. \ref{fig_overall} presents the detailed neural network architecture of our proposed recognition model, explicitly showing the specific structure and connections of the PRN, NAN, and ARN.

\begin{table}[t]
\centering
\caption{Description of networks and their specific roles within NAEL}
\includegraphics[width= 0.48 \textwidth]{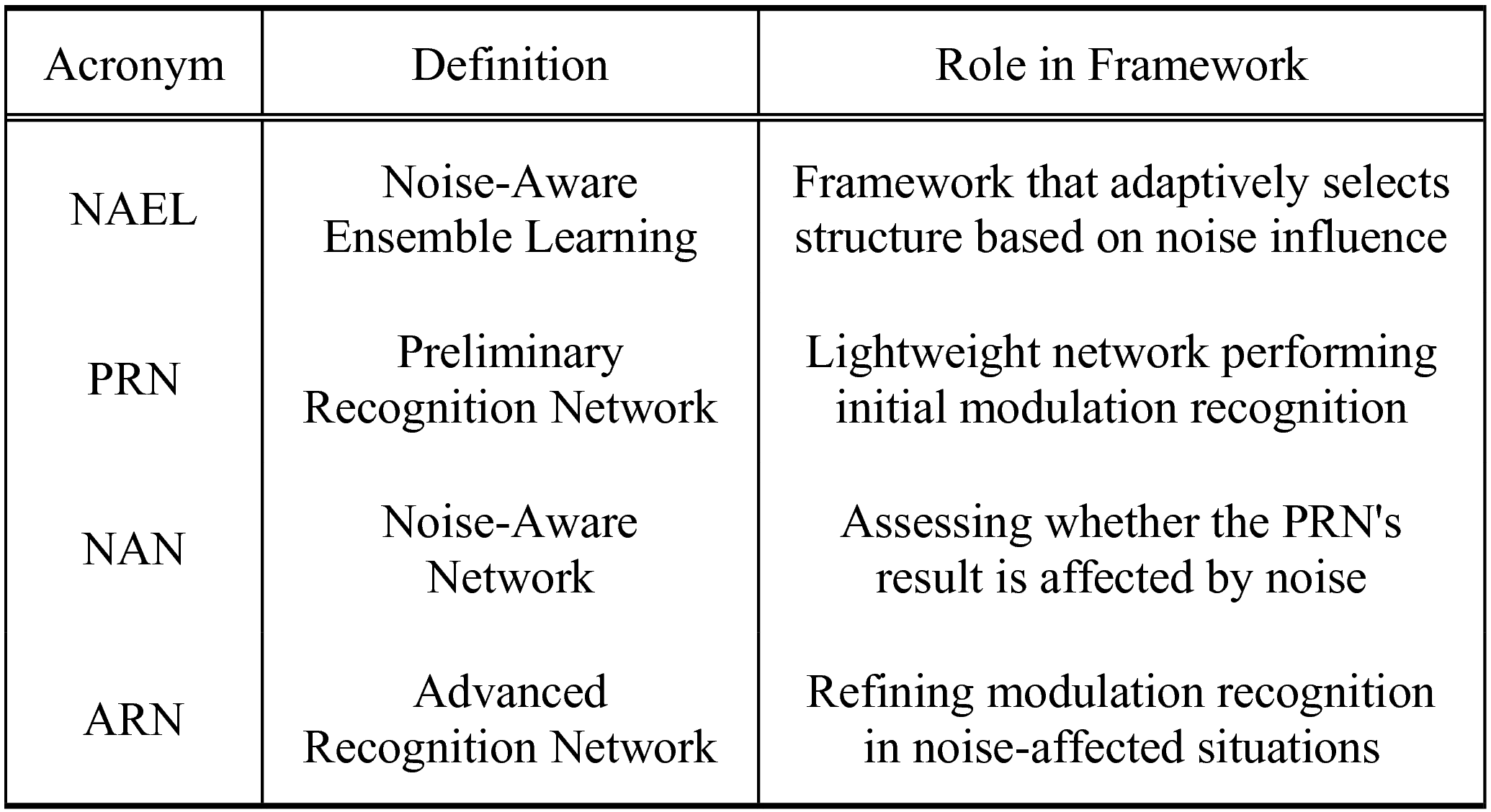}
\label{table_acronym}
\end{table}

\subsection{FE Block}
The PRN and ARN within the NAEL framework utilize FE blocks that are both powerful and computationally efficient, unlike techniques used in conventional CNNs, where spatial features are extracted from the input feature map using standard convolution (SC) layers. The output feature map of a SC operation \(F_{os}\) is calculated as:
\begin{equation}
F_{os}(h,w,c_{o})=\sum_{i,j,c_{i}} K_{s}(i,j,c_{i},c_{o}) \cdot F_{i}(h+i-1,w+j-1,c_{i})
\label{SC_output}
\end{equation}
where \( K_{s} \) represents the SC kernel and \( F_{i} \) is the input feature map. The dimensions of \( F_{i} \), \( F_{os} \), and \( K_{s} \) are \( H_{f} \times W_{f} \times C_i \), \( H_{f} \times W_{f} \times C_o \), and \( H_{k} \times W_{k} \times C_i \times C_o \), respectively. Here, \( H_{f} \) represents the height of the feature map, \( W_{f} \) the width of the feature map, \( C_i \) the number of input feature map channels, \( C_o \) the number of output feature map channels, \( H_{k} \) the height of the kernel, and \( W_{k} \) the width of the kernel. By padding the input feature map with zeros and setting the stride to 1, the convolution operation moves the kernel one step at a time, resulting in an output feature map (\( F_{os} \)) of the same height and width as the input feature map (\( F_{i} \)). As shown in (\ref{SC_output}), the SC layer applies a distinct filter to each input feature map channel. 
The computational cost of a SC operation is expressed as
\begin{equation}
H_{k} \cdot W_{k} \cdot C_{i} \cdot C_{o} \cdot H_{f} \cdot W_{f}.
\label{SC_cost}
\end{equation}

To reduce the computational cost of extracting spatial features, we decomposed the SC operation into two steps: depth-wise and point-wise convolution operations. Unlike standard convolution, which uses multiple filters (one for each output channel) that span all input channels, depth-wise convolution employs a single filter per input channel. The feature-map output of the depth-wise convolution operation \(F_{od}\) is expressed as follows:
\begin{equation}
F_{od}(h,w,c_{i})=\sum_{i,j} K_{d}(i,j,c_{i}) \cdot F_{i}(h+i-1,w+j-1,c_{i})
\label{DC_output}
\end{equation}
where \( K_{d} \) denotes the depth-wise convolution kernel with dimensions \( H_{k} \times W_{k} \times C_i \). Each channel operation in (\ref{DC_output}) uses a corresponding filter from \( K_{d} \), generating an output channel in \( F_{od} \). Thus, the number of channels in \( F_{i} \) and \( F_{od} \) remains the same. Following depth-wise convolution, a point-wise convolution with a 1\(\times\)1 kernel combines the channels of \( F_{od} \) to produce \( C_{o} \) output channels. The total computational complexity of depth-wise and point-wise convolutions is expressed as follows:
\begin{equation}
H_{k} \cdot W_{k} \cdot C_{i} \cdot H_{f} \cdot W_{f} + C_{i} \cdot C_{o} \cdot H_{f} \cdot W_{f}
\label{DC_cost}
\end{equation}
where the first term represents the computational cost of depth-wise convolution, and the second term represents the cost of point-wise convolution. This method of factorizing SC into two operations significantly reduces computational complexity while maintaining high spatial feature extraction performance\cite{MobileNetV1,DSconv}.

\begin{figure}[t]
\centering
\includegraphics[width= 0.48 \textwidth]{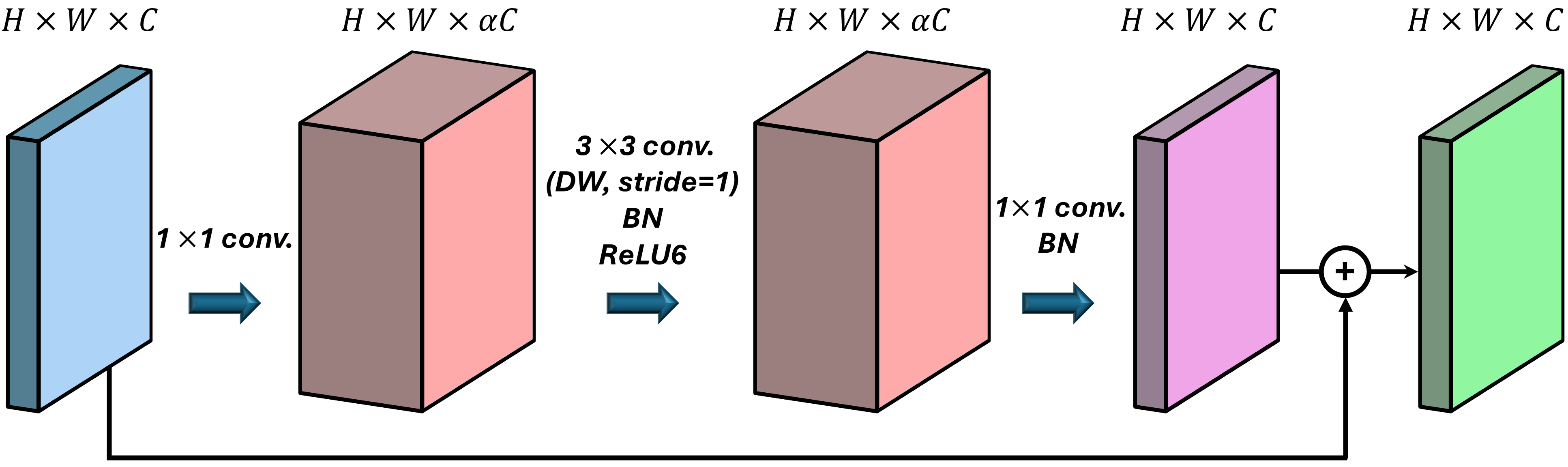}
\caption{Block diagram illustrating the structure used in the FE block, excluding the first stage.}
\label{fig_FEBlock}
\end{figure}

\begin{figure}[t]
\centering
\includegraphics[width= 0.48 \textwidth]{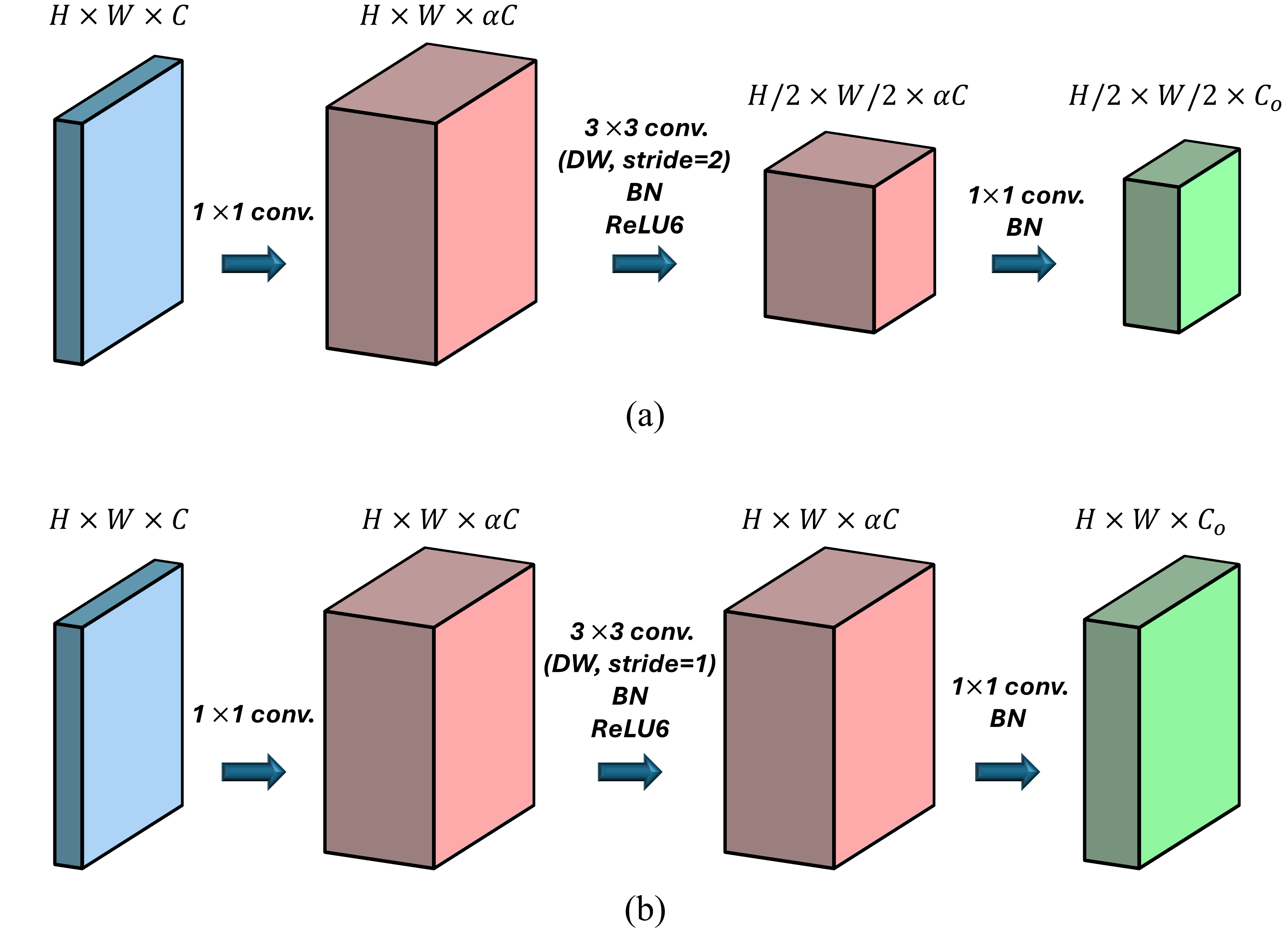}
\caption{Block diagrams illustrating the structures corresponding to the first stage of the FE block. Structure performing (a) both channel adjustment and downsampling operation, (b) only channel adjustment.}
\label{fig_FEBlock2}
\end{figure}

Our proposed recognition neural network employs FE blocks following the structures illustrated in Figs. \ref{fig_FEBlock} and \ref{fig_FEBlock2}. The structure depicted in Fig. \ref{fig_FEBlock} is used repetitively from the second stage onward, while the structure in Fig. \ref{fig_FEBlock2} is used only in the first stage. In the first stage, depending on the parameters used in the FE block design, the model can adjust the number of output channels, perform downsampling, or both simultaneously.

As shown in Fig. \ref{fig_FEBlock}, a feature map input into the FE block undergoes channel expansion by a factor \(\alpha\) through a 1\(\times\)1 point-wise convolution. Subsequently, spatial features were extracted using depth-wise convolution. Batch normalization (BN) is then applied to enhance training efficiency\cite{BN}. An activation function was employed to incorporate nonlinearity, enabling the neural network to learn complex patterns. Typically, CNNs use the ReLU activation function\cite{ReLU} due to its simplicity and computational efficiency. We used a variant called ReLU6, which limits activations to the range of 0 to 6, thereby enhancing network robustness \cite{MobileNetV1}. After applying the activation function, a 1\(\times\)1 point-wise convolution layer combined the outputs of the depth-wise convolution. The number of output channels from this convolution layer is adjusted to match the input feature map dimensions. Finally, a skip connection adds the initial input feature map to the output, producing the final output feature map. An FE block with this structure has the following advantages:
\begin{enumerate}
\item {By expanding channels for spatial feature extraction and subsequently compressing them, this method reduces parameter usage and computational complexity.}
\item{Introducing skip connections between the input and output feature maps helps mitigate learning difficulties that can arise with deep network architectures.}
\item{The FE block processes inputs and outputs with a reduced number of channels. This approach contrasts with traditional CNN structures \cite{ResNet}, which typically employ wide-channel feature maps for skip connections. Using fewer channels for skip connections significantly reduces memory usage in our model.}
\end{enumerate}

Downsampling operations can be integrated into the first stage of the FE block to enhance computational efficiency and extract key features effectively. If the block parameters are set to include a downsampling operation, the structure illustrated in Fig. \ref{fig_FEBlock2}(a) is used. This structure reduces the spatial dimensions of the output feature map by half and adjusts the number of output channels to \(C_{o}\). Otherwise, the structure illustrated in Fig. \ref{fig_FEBlock2}(b) is used. Here, the FE block retains the original spatial dimensions of the input feature map while adjusting the number of output channels to \(C_{o}\).

This FE block structure is particularly advantageous for radar modulation recognition due to its ability to efficiently extract time-frequency features while maintaining computational efficiency. By employing depth-wise convolution, the model effectively captures spatial dependencies within individual feature map channels. The subsequent point-wise convolution integrates the extracted features across channels, enabling the model to learn complex radar modulation patterns more effectively. Additionally, the incorporation of skip connections helps preserve critical modulation characteristics, ensuring robust recognition performance even in low-SNR conditions. With these design choices, the proposed framework can effectively identify modulation schemes while balancing accuracy, efficiency, and robustness.

\subsection{Modulation Recognition Neural Networks}
This subsection provides a detailed explanation of the structures of PRN and ARN that constitute the proposed recognition model. PRN and ARN utilize FE blocks for modulation recognition. Table \ref{table_network} summarizes the input dimensions and parameters used for each block, where \( C_{o} \) denotes the number of channels in the block output, \(\alpha\) implies channel expansion factor, \( r \) represents the number of repetitions for spatial feature extraction, and \( s \) indicates the convolution layer stride in the first stage of the block. Based on our experimental results, setting the parameters listed in Table \ref{table_network} yielded the best modulation recognition performance. For the FE block, the structure shown in Fig. \ref{fig_FEBlock2} is used in the first stage, and the structure in Fig. \ref{fig_FEBlock} is repeated \(r-1\) times for subsequent stages. In addition, if \(s=2\), the structures shown in Fig. \ref{fig_FEBlock2}(a) is used in the first stage; if \(s=1\), the structure in Fig. \ref{fig_FEBlock2}(b) is used.

\begin{table}[t]
\centering
\caption{Detailed structure of the recognition networks within the NAEL framework}
\includegraphics[width= 0.48 \textwidth]{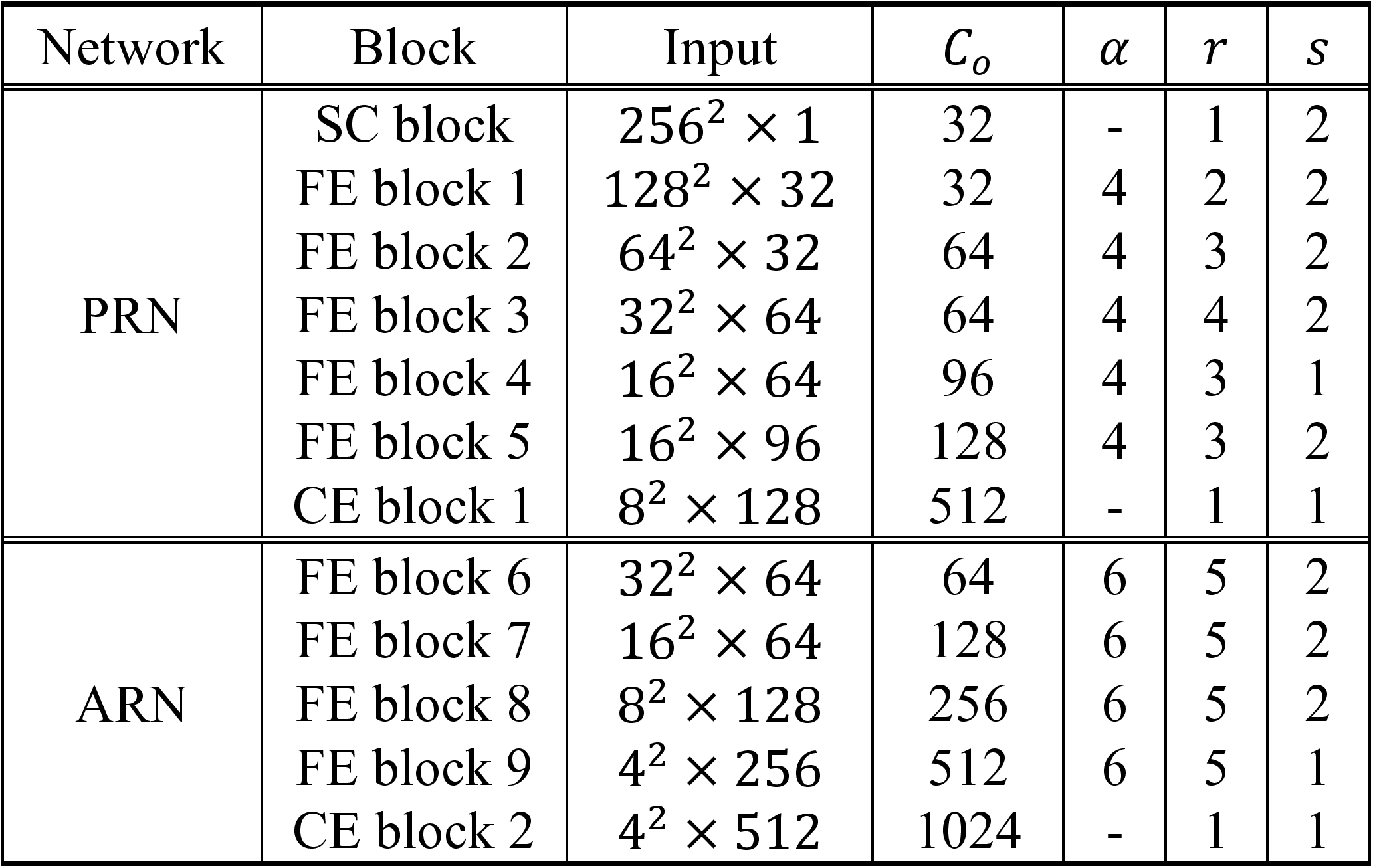}
\label{table_network}
\end{table}

As shown in Fig. \ref{fig_overall}, PRN initially extracts features from the input TFI using a SC block. The first neural network process involved extracting significant low-level features. In the initial stage of the PRN, we used a standard 3\(\times\)3 convolution layer, followed by BN and ReLU6 activation functions. This SC block effectively captures basic features from the input data.
We then utilized FE blocks based on depth-wise convolution in the subsequent stages.

For classification, transforming a feature map into a one-dimensional feature vector is crucial for effective processing by a classifier based on fully connected layers. To achieve this, we employed a channel expansion (CE) block after the last FE block. This block consists of a 1\(\times\)1 convolution layer, followed by BN and ReLU6 activation. The CE block enhances the representation of each channel in \(\mathcal{F}\), the output of the last FE block, to capture essential information. 
Subsequently, a one-dimensional feature vector is produced by averaging the values across channels of the CE block output. This feature vector is then fed into a classifier composed of fully connected layers followed by a softmax layer. The classifier computes class scores based on the feature vector, identifying the modulation scheme with the highest score.

In cases where the NAN determines significant noise influence on PRN's recognition results (as explained in a subsequent subsection), ARN is activated for more precise analysis. Although both ARN and PRN utilize FE blocks for spatial feature extraction, ARN differs from PRN in the following ways:

\begin{enumerate}
\item {The ARN employs FE blocks with a higher channel expansion ratio \(\alpha\) compared to the PRN, facilitating more detailed feature extraction.}
\item{Through the extraction of higher-level features, ARN generates final feature maps that are reduced in spatial dimensions but increased in channel depth compared to those of the PRN.}
\item{As illustrated in Fig. \ref{fig_overall}, some of the inputs to ARN's FE blocks include outputs from the PRN block. This data-reuse strategy enhances the efficiency of the modulation recognition model design.}
\end{enumerate}

\begin{figure*}[t]
\centering
\includegraphics[width=\textwidth]{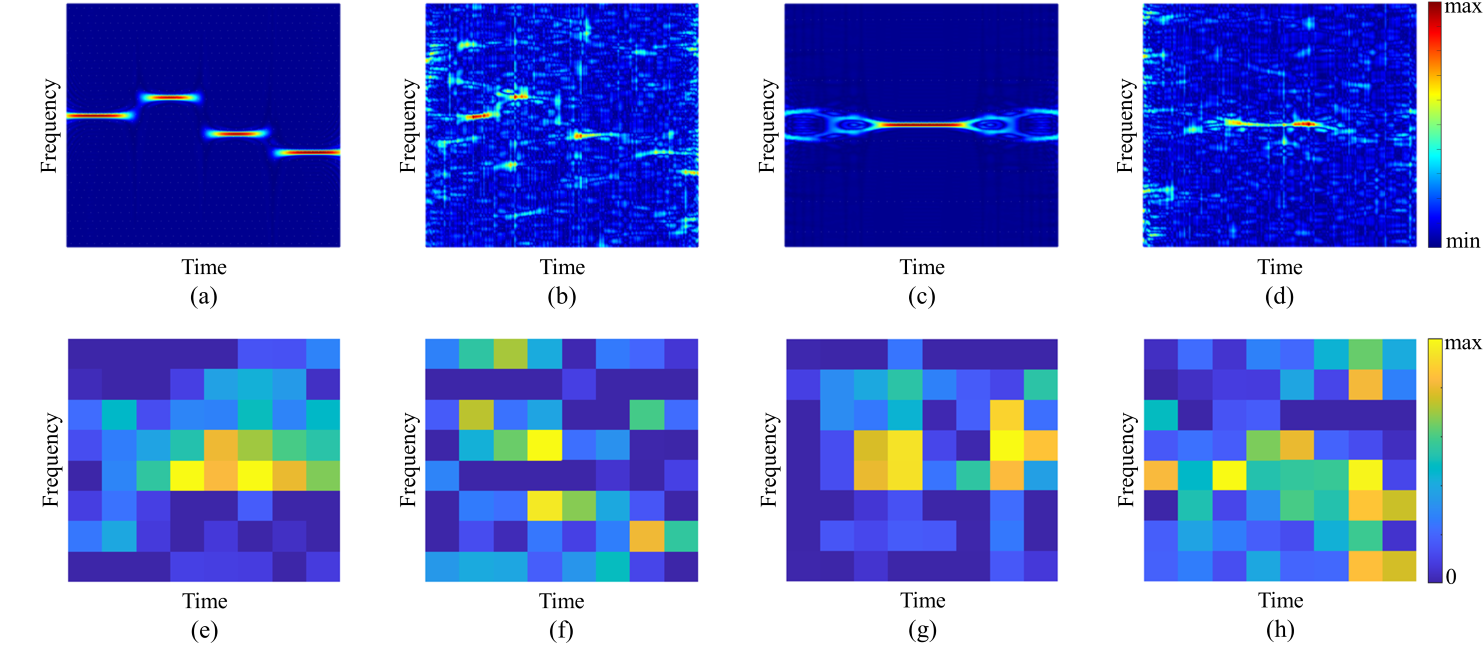}
\caption{Examples of TFIs and gradient maps for Costas and T2 code simulated signals; (a) and (c) TFIs of transmitted Costas and T2 code signals, (b) and (d) TFIs of received Costas and T2 code signals, (e) and (g) gradient maps for the correctly classified cases of Costas and T2 codes, (f) and (h) gradient maps for the misclassified cases of Costas and T2 codes.}
\label{fig_CAMs}
\end{figure*}

\subsection{Noise-Aware Neural Network}
A key proposal of this study involves using a noise-aware method to determine the impact of noise on recognition results derived by the PRN. To achieve this, the NAN generates a gradient map illustrating the regions of the input TFI on which the PRN focuses to derive modulation recognition results. The method for generating the gradient map is based on computing gradient information from the PRN's feature map \(\mathcal{F}\), as gradient information is crucial for visually highlighting the data influencing the decision \cite{GRADCAM}. To obtain the gradient map, the process starts by computing the gradient between the \(c\)-th class score \(y^{c}\) (the highest among computed class scores) from the PRN's classifier output, and the \(k\)-th channel of the feature map \(\mathcal{F}^{k}\) from the output of the PRN's CE block. This gradient \(\partial y^{c}/ \partial \mathcal{F}^{k}\) represents the influence of \(\mathcal{F}^{k}\) on \(y^{c}\). To determine the importance of the \(k\)-th feature map channel for the \(c\)-th class score, an importance weight \(w^{c}(k)\) is introduced. This weight quantifies how significantly the \(k\)-th channel contributes to recognizing the \(c\)-th modulation class and is computed using global average pooling across the width and height dimensions of \(\mathcal{F}^{k}\). This process is summarized as follows:
\begin{equation}
w^{c}(k)=\frac{1}{{H \times W}} \sum_{j=1}^{H} \sum_{i=1}^{W} \frac{\partial y^{c}}{\partial \mathcal{F}^{k}(t_{i},f_{j})}   
\label{weight_calculation}
\end{equation}
where \(H\) and \(W\) denote the height and width of \(\mathcal{F}\), and \(\mathcal{F}^{k}(t_{i},f_{j})\) are the feature map values for the \(i\)-th height and \(j\)-th width at the \(k\)-th channel.

By performing a weighted sum on \(\mathcal{F}^{k}\) with the importance weights and applying the ReLU activation function, a gradient map \(\mathcal{G}^{c}\) is produced. The gradient map \(\mathcal{G}^{c}\) visually highlights the time-frequency regions in the input TFI that significantly contribute to the PRN's recognition of the \(c\)-th modulation scheme. The gradient map is computed as follows:
\begin{equation}
\mathcal{G}^{c}(t,f)=\textup{ReLU} \left ( \mathit{\sum_{k}^{} \mathcal{F}^{k}(t,f) \cdot w^{c}(k)    }  \right ).
\label{map_generation}
\end{equation}

Gradient maps can serve as visualizations of the neural network's decision-making process, allowing for the diagnosis of faults using this explainability \cite{CAMApp}. To explicitly evaluate the influence of noise on recognition, the proposed method analyzes the gradient map \(\mathcal{G}^{c}\). Specifically, our proposed method exploits the property that wireless signals occupy regions near the center frequency when analyzed in the frequency domain. Therefore, if high gradient values appear predominantly in noise-dominant regions rather than around the signal’s center frequency, it indicates that the recognition process can be significantly influenced by noise. This relationship arises because the distribution of the gradient map is directly influenced by the SNR. As the SNR increases, the network can focus more accurately on the signal components, resulting in a gradient map with high energy concentrated at the actual signal locations (i.e., regions around the center frequency). Conversely, when the SNR decreases, noise can be misinterpreted as signal features, leading to energy spreading in the gradient map beyond the true signal regions. This phenomenon introduces ambiguity in the recognition process, making it crucial for the NAN to assess the reliability of the recognition results based on gradient map characteristics.

Fig. \ref{fig_CAMs} shows examples of gradient maps for simulated signals in cases of correct classification and misclassification. Costas and T2 code signals were used in an environment with an SNR of \(-11\) dB. In the proposed recognition model, the feature map \(\mathcal{F}\) has dimensions of 8\(\times\)8, resulting in a gradient map of the same 8\(\times\)8 size. As shown in Fig. \ref{fig_CAMs}(e) and (g), during correct classification, the gradient map exhibits the highest values at indices 4 and 5 in the frequency domain. The ES system was assumed to accurately estimate the center frequency for all simulated signals before performing modulation recognition, with a frequency offset applied to align the center frequency with the middle of the TFI frequency axis. Therefore, for correctly classified cases, the highest gradient values are concentrated at indices 4 and 5, which correspond to the signal's center frequency. In the case of misclassification, as shown in Fig. \ref{fig_CAMs}(f) and (h), high values appear at positions other than the center frequency. This occurs because the gradient map highlights noisy areas where the signal does not exist, leading the PRN to focus on noise and make incorrect decisions.

\begin{figure}[t]
\centering
\includegraphics[width= 0.48 \textwidth]{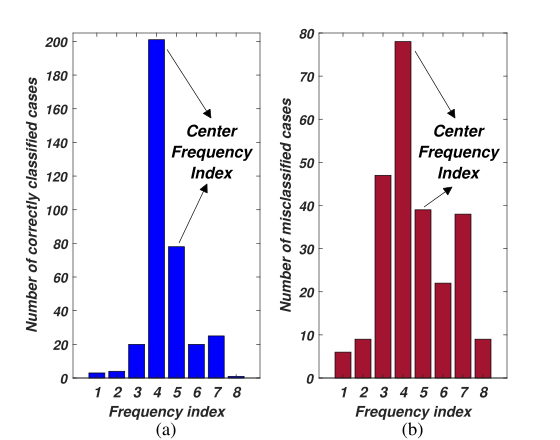}
\caption{Distribution of the frequency index \(f_{max}\) in (a) correct and (b) incorrect classifications at an SNR of \(-15\) dB.}
\label{fig_histogram}
\end{figure}

Fig. \ref{fig_histogram} shows the distribution of \(f_{max}\) for both correctly and incorrectly classified cases. For this analysis, 50 simulated signals for each of the 12 modulation schemes were generated in an environment with an SNR of \(-15\) dB. The frequency index \(f_{max}\) represents the frequency location where the gradient map \(\mathcal{G}^{c}(t,f)\) has the highest activation, indicating the frequency region most influential for modulation recognition. The \(f_{max}\) can be obtained as follows:

\begin{equation}
f_{max}= \textup{arg}\max_{f} \left ( \sum_{t=1}^{W} \mathcal{G}^{c}(t,f) \right ).
\label{fmax_calculation}
\end{equation}

As seen in Fig. \ref{fig_histogram}(a), which shows the distribution of \(f_{max}\) when correctly classified, the highest values are concentrated at frequency indices 4 and 5, corresponding to the center frequency. In contrast, Fig. \ref{fig_histogram}(b) illustrates the distribution in cases of misclassification, revealing a noticeably higher number of outliers compared to when classifications are correct. This difference highlights a clear distinction between correct and incorrect recognition results. These findings strongly indicate that utilizing a gradient map is highly effective in evaluating the reliability of recognition outcomes, as it helps detect anomalies associated with misclassification.

The NAN framework quantifies noise impact by analyzing the distribution of activation in the gradient map. Specifically, when the PRN correctly classifies a signal, the highest activation values in the gradient map \(\mathcal{G}^{c}(t,f)\) tend to be concentrated at the center frequency, indicating that the model focuses on meaningful signal features. However, in cases of misclassification, these activations are more dispersed, suggesting that the model is influenced by noise rather than the actual signal. To systematically assess this effect, we measure the spatial distribution of high-activation areas within the gradient map and compare them with expected signal characteristics.

Fig. \ref{fig_histogram} demonstrates the usefulness of the gradient map using \(f_{max}\); however, a deep learning-based classifier that utilizes all elements of the gradient map as input can achieve better noise awareness performance. While using \(f_{max}\) as a decision criterion provides a straightforward approach to assessing noise influence, it only considers the frequency index with the highest activation in the gradient map. This approach does not fully capture the spatial distribution of activations across the entire time-frequency representation. In contrast, by utilizing the entire gradient map as input, the proposed classifier can learn more complex patterns associated with noise influence, leading to improved classification performance in diverse noise environments. Therefore, the NAN framework further refines noise assessment by using a deep learning-based classifier. Instead of relying solely on a single frequency index, the classifier takes the entire gradient map as input, allowing it to learn more complex patterns of noise interference.

The gradient map is first flattened into a one-dimensional feature vector with 64 elements. It is then processed through two fully connected layers, which sequentially expand the feature dimension to 256 and then to 512. Each layer applies batch normalization and a ReLU6 activation to improve learning stability. Finally, a fully connected layer maps the 512-dimensional feature vector to two output neurons, representing the classification outcomes: reliable (not noise-affected) and unreliable (noise-affected). A softmax activation function is applied to the output layer to convert the final classification scores into probability values, ensuring that the model assigns a confidence score to each possible outcome. If the classifier assigns a higher probability to the reliable category, the PRN’s recognition result is accepted as the final decision. If the unreliable category receives a higher probability, ARN is activated to conduct a more advanced feature extraction process and refine the recognition result.

The use of fully connected layers ensures that the classifier does not rely solely on a single frequency index but instead learns to extract discriminative features from the entire gradient map. Additionally, the softmax classifier assigns probabilistic confidence scores to the predictions, allowing the system to dynamically adapt to varying SNR conditions. Unlike fixed-threshold methods, which may not generalize well across different noise environments, the learned model effectively captures non-linear dependencies, resulting in a more robust noise-aware classification. This classification process inherently accounts for the impact of varying SNR levels as the network learns to recognize the structural differences in gradient maps across different SNR environments. By analyzing how the gradient map's energy distribution shifts under different SNR conditions, the NAN can dynamically adapt its noise-aware decision-making, leading to improved recognition robustness even in challenging scenarios. 

\begin{table}[t]
\caption{Modulation Parameters for Dataset Generation}
\centering
\begin{tabular}{|c|c|c|}
\hline
Scheme            & Param. & Value                  \\ \hline
\multirow{1}{*}{All schemes}       & SNR (dB)       & \(U(-15, 5)\)        \\ \hline
\multirow{2}{*}{LFM}       & \(f_{c}\)       & \(U(1/8f_{s}, 1/4f_{s})\)        \\
& \(B\)           & \(U(1/20f_{s}, 1/8f_{s})\)  \\ \hline
\multirow{3}{*}{Costas}    & \(f_{min}\)     & \(U(1/40f_{s}, 1/10f_{s})\) \\
& \(L_{hs}\)      & \{4, 6\}                       \\
& \(\quad f_{hop} \quad\)     & \(U(1/40f_{s}, 3/40f_{s})\) \\ \hline
\multirow{3}{*}{Barker}    & \(f_{c}\)       & \(U(1/8f_{s}, 1/4f_{s})\)   \\ 
& \(L_{B}\)       & \{7, 11, 13\}                   \\ 
& \(N_{sc}\)         & \{20, 24, 28, 32\}               \\ \hline
\multirow{3}{*}{Frank, P1} & \(f_{c}\)       & \(U(1/8f_{s}, 1/4f_{s})\)   \\ 
& \(N_{sc}\)         & \{5, 6, 7\}                     \\  
& \(M\)           & \{6, 7, 8\}                     \\ \hline
\multirow{3}{*}{P2}        & \(f_{c}\)       & \(U(1/8f_{s}, 1/4f_{s})\)   \\ 
& \(N_{sc}\)         & \{5, 6, 7\}                     \\  
& \(M\)           & \{6, 8\}                       \\ \hline
\multirow{3}{*}{P3, P4}    & \(f_{c}\)       & \(U(1/8f_{s}, 1/4f_{s})\)   \\ 
& \(N_{sc}\)         & \{5, 6, 7\}                     \\  
& \(N_{c}\)        & \{36, 49, 64\}                  \\ \hline
\multirow{2}{*}{T1, T2}    & \(f_{c}\)       & \(U(1/8f_{s}, 1/4f_{s})\)  \\  
& \(k\)       & \{5, 6, 7\}                     \\ \hline
\multirow{2}{*}{T3, T4}    & \(f_{c}\)       & \(U(1/8f_{s}, 1/4f_{s})\)   \\ 
& \(\Delta F\)    & \(\qquad U(1/20f_{s}, 1/10f_{s}) \qquad\) \\ \hline
\end{tabular}
\label{table_parameter}
\end{table}

\section{Performance Analysis}
This section presents the modulation recognition performance of the classification model based on the NAEL framework. Additionally, we provide a comparative performance analysis of the proposed and existing classification models. For the performance analysis, a training dataset was generated in a simulation environment, and a test dataset was produced using real measured signals.

\subsection{Training Recognition Model}
A training dataset was built to train the recognition model, generating 5,000 simulated signals for each of the 12 modulation schemes. Table \ref{table_parameter} lists the parameters used in the signal generation, where \(f_{s}\), \(f_{min}\), \(L_{hs}\), \(f_{hop}\), \(L_{B}\), \(N_{sc}\), \(k\), and \(\Delta F\) represent sampling frequency, fundamental frequency of the Costas code, hopping sequence length of the Costas code, frequency spacing of the Costas code, length of the Barker code, the number of samples per carrier frequency cycle, the number of segments, and modulation bandwidth, respectively. Additionally, \(U(\cdot)\) denotes a uniform distribution.

The proposed NAEL framework consists of three neural networks, each trained individually whereas keeping the weights of the other networks constant during training. For the performance analysis, all classification models used in this study, including the NAEL-based model, were trained using the Adam optimizer \cite{ADAM}. The training was conducted for 100 epochs, resulting in 187,500 weight updates. The input data fed into all neural networks were normalized to have a mean of 0 and a variance of 1. All experiments were conducted on a workstation equipped with an AMD Ryzen 9 5950X CPU and an NVIDIA GeForce RTX 3080 Ti GPU, using the PyTorch framework.

\begin{figure}[t]
\centering
\includegraphics[width= 0.48 \textwidth]{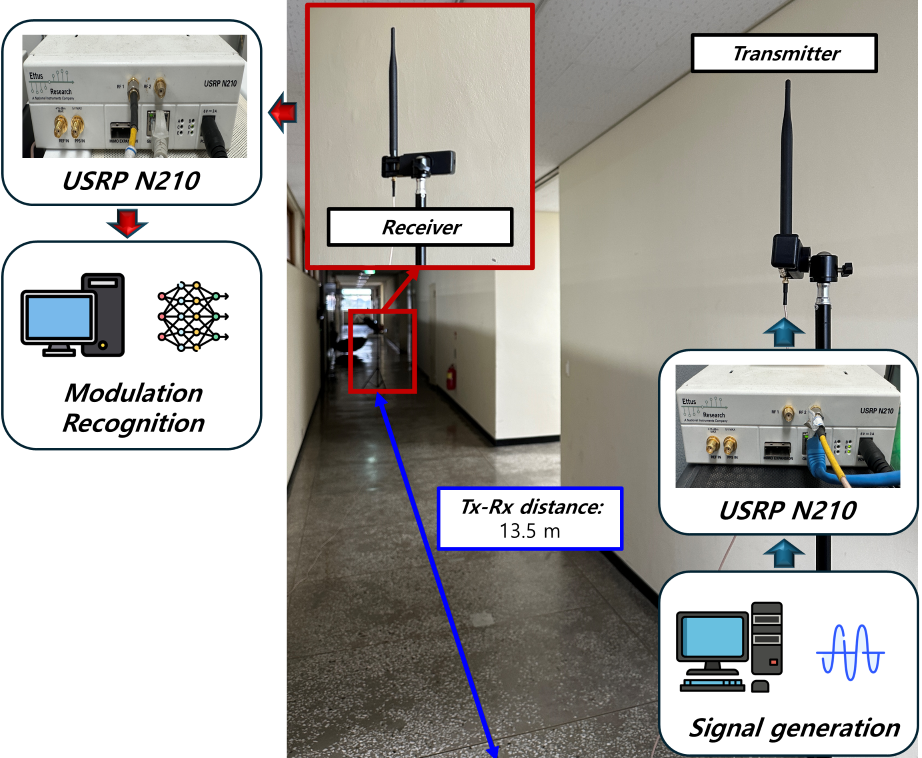}
\caption{Illustration of measurement environment for test dataset generation under realistic conditions.}
\label{fig_experiment}
\end{figure}

\begin{figure*}[t]
\centering
\includegraphics[width=\textwidth]{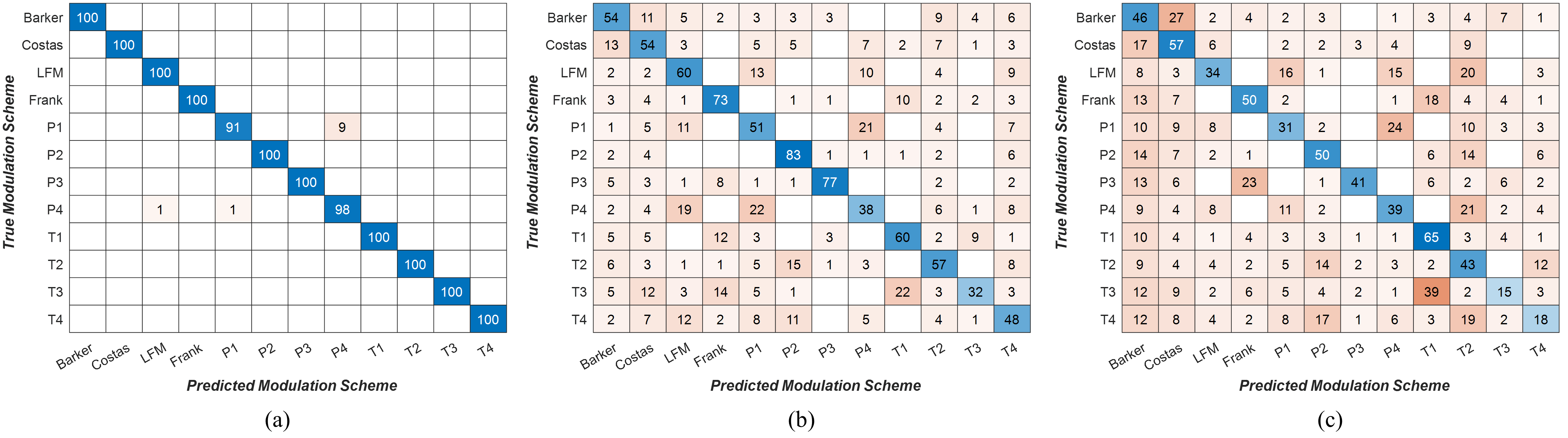}
\caption{Confusion matrix of the proposed recognition model using the test dataset collected in the (a) first scenario (high-SNR case), (b) second scenario (low-SNR case), and (c) third scenario (very low-SNR case).}
\label{fig_confusionmatrix}
\end{figure*}

\subsection{Test Dataset Generation}
The test dataset for performance analysis was constructed by collecting signals in a real-world environment, as shown in Fig. \ref{fig_experiment}. In this experiment, universal software radio peripheral (USRP) N210 software-defined radios (SDRs)\cite{USRP} were employed. SDRs provide flexible and programmable platforms capable of transmitting and receiving various signal types, making them suitable for diverse wireless applications such as radar modulation recognition. Although practical SDR devices like the USRP N210 typically have constraints such as limited sampling rates and potential hardware-induced noise or distortion, their versatility and ease of use still make them well-suited for radar modulation recognition research.

In our experiment, radar signals with 12 modulation schemes were generated using MATLAB. The generated digital signals were converted to analog signals using the USRP N210 SDR platform. The converted analog signals were transmitted via an antenna connected to a transmitting USRP N210 and received by an antenna connected to a distant receiving USRP N210. Vertical antennas were used for both transmission and reception, positioned 13.5 m apart. The radar signals used in this experiment had a center frequency of 1,902.5 MHz. The collected signals were digitized at a sampling rate of 10 MHz using a USRP N210 and imported into a computer for modulation recognition.

These collected signals inherently reflect real-world conditions, incorporating multiple factors that deviate from an idealized additive white Gaussian noise (AWGN)-only environment. As illustrated in Fig. \ref{fig_experiment}, multipath interference occurs due to the presence of obstacles, causing signal reflections that alter channel characteristics. Additionally, the use of USRP N210 hardware introduces hardware-induced distortions such as phase noise and frequency offset, further differentiating the experimental setting from a controlled simulation. By capturing these diverse real-world effects, the test dataset enables a comprehensive evaluation of the robustness of the proposed method.

The signals were recorded for three different scenarios by varying the transmission power to analyze performance under various SNR conditions. In each scenario, 100 signals were collected for each of the 12 modulation schemes. The test dataset comprised three scenarios: one with a high-SNR condition and two with low- and very low- SNR conditions. The estimated SNRs for the first, second, and third scenarios were approximately \(-4\), \(-15\), and \(-17\) dB, respectively. Notably, in the experiments using the USRP, the exact SNR for each scenario could not be calculated. However, an approximate SNR can be estimated by assuming that the power of the collected signal \(y(n)\) is the sum of the power of the signal \(x(n)\) and that of the signal collected in an environment without transmitted signals (i.e., received data with only noise \(w(n)\) presence).

\subsection{Analysis Results on Experimental Data}
In this subsection, we present the results of the performance analysis of the proposed model using the test dataset. Additionally, we compared the modulation recognition performance of the proposed model with existing models known for their high accuracy and efficiency despite their low computational complexity.

Fig. \ref{fig_confusionmatrix} shows the confusion matrix of the proposed model for the three different scenarios. A confusion matrix is a widely used metric that visually represents the difference between true labels and predicted results in classification problems. In the confusion matrix, the columns represent the actual transmitted modulation schemes, and the rows represent the predicted modulation schemes. The modulation recognition results were counted in the corresponding grids, and the total sum of all grids was 1,200 since 100 signals were collected for each modulation scheme. 

\begin{table}[t]
    \caption{PCC and FLOPs of Various Recognition Models Using the Test Dataset Collected in the First Scenario}
    \centering
    \setlength{\tabcolsep}{17pt} 
    \renewcommand{\arraystretch}{1.1} 
    \begin{tabular}{lcc}
        \toprule
        Recognition Model & PCC (\%) & MFLOPs \\
        \midrule
        LPI-Net & 93.8 & 69 \\
        LWCNN & 98.0 & 507 \\
        MobileNetV1 & 98.8 & 701 \\
        MobileNetV2 & 98.7 & 416 \\
        ShuffleNetV1 2.0x & 98.8 & 692 \\
        ShuffleNetV2 2.0x & 98.8 & 766 \\
        MNasNet 1.3 & 98.6 & 702 \\
        Vision Transformer & 98.8 & 5,540 \\
        \textbf{NAEL} & \textbf{99.1} & \textbf{402} \\
        \bottomrule
    \end{tabular}
    \label{scn1}
\end{table}

\begin{table}[t]
    \caption{PCC and FLOPs of Various Recognition Models Using the Test Dataset Collected in the Second Scenario}
    \centering
    \setlength{\tabcolsep}{17pt} 
    \renewcommand{\arraystretch}{1.1} 
    \begin{tabular}{lcc}
        \toprule
        Recognition Model & PCC (\%) & MFLOPs \\
        \midrule
        LPI-Net & 24.8 & 69 \\
        LWCNN & 43.4 & 507 \\
        MobileNetV1 & 49.5 & 701 \\
        MobileNetV2 & 51.1 & 416 \\
        ShuffleNetV1 2.0x & 51.6 & 692 \\
        ShuffleNetV2 2.0x & 52.1 & 766 \\
        MNasNet 1.3 & 54.5 & 702 \\
        Vision Transformer & 51.4 & 5,540 \\
        \textbf{NAEL} & \textbf{55.4} & \textbf{606} \\
        \bottomrule
    \end{tabular}
    \label{scn2}
\end{table}

\begin{table}[t]
    \caption{PCC and FLOPs of Various Recognition Models Using the Test Dataset Collected in the Third Scenario}
    \centering
    \setlength{\tabcolsep}{17pt} 
    \renewcommand{\arraystretch}{1.1} 
    \begin{tabular}{lcc}
        \toprule
        Recognition Model & PCC (\%) & MFLOPs \\
        \midrule
        LPI-Net & 23.6 & 69 \\
        LWCNN & 30.2 & 507 \\
        MobileNetV1 & 34.8 & 701 \\
        MobileNetV2 & 33.4 & 416 \\
        ShuffleNetV1 2.0x & 35.5 & 692 \\
        ShuffleNetV2 2.0x & 33.9 & 766 \\
        MNasNet 1.3 & 35.9 & 702 \\
        Vision Transformer & 33.8 & 5,540 \\
        \textbf{NAEL} & \textbf{38.3} & \textbf{660} \\
        \bottomrule
    \end{tabular}
    \label{scn3}
\end{table}

In the first scenario (as seen in Fig. \ref{fig_confusionmatrix}(a)), where the signal power is relatively high, the proposed recognition model exhibits low confusion, resulting in high classification performance. However, as shown in Fig. \ref{fig_confusionmatrix}(b), many misclassifications occur as the SNR decreases due to the increasing influence of noise on the signal characteristics. In particular, distinguishing between Frank and P3, as well as P1 and P4, becomes significantly challenging. This is because these modulation schemes rely heavily on precise phase transitions for differentiation, and in low SNR conditions, phase information is highly susceptible to noise-induced distortions\cite{EuRAD}. As a result, the model may incorrectly classify these modulation types due to the increased similarity in their observed signal patterns. Additionally, when the signal power was extremely low, the classification performance for all modulation schemes decreased, as shown in Fig. \ref{fig_confusionmatrix}(c). This degradation occurs because the noise level highly exceeds the signal power, making it difficult for the model to extract meaningful features from the received data. Consequently, the overall classification accuracy decreases, with more modulation schemes being misclassified due to indistinguishable modulation characteristics in a high-noise environment.

Table \ref{scn1}, \ref{scn2}, and \ref{scn3} present the comparison results between the proposed model and existing classification models across the three scenarios. We used LPI-Net\cite{CNNCWD}, LWCNN\cite{LWCNN}, MNasNet\cite{MNasNet}, ShuffleNetV1\cite{ShuffleNetV1}, ShuffleNetV2\cite{ShuffleNetV2}, MobileNetV1\cite{MobileNetV1}, MobileNetV2\cite{MobileNetV2}, and Vision Transformer\cite{VITori} for comparison. These models are known for their high accuracy and efficiency, as described in Section \Romannum{1}. To evaluate the efficiency of each model, we used the probability of correct classification (PCC) versus mega floating point operations (MFLOPs). The computational load of the proposed recognition model varied with each inference; therefore, the average FLOPs were calculated for all 1,200 inferences.

In the high SNR scenario shown in Table \ref{scn1}, the proposed model exhibited the highest recognition performance and the lowest computational cost compared to the other models. At high SNR levels, most signals are classified correctly using the PRN alone, which results in low computational cost while maintaining high PCC. Therefore, small variations in FLOPs occur due to occasional ARN activation when the model detects uncertain cases, leading to minor fluctuations in PCC. This is because the NAN determined that most of the input data were not significantly affected by noise. Consequently, recognition could be performed using only the PRN at a relatively low computational cost, without activating the ARN. In fact, the NAN of our proposed model activated the ARN 20 times out of 1,200 predictions in the first scenario, leading to an average computational load of approximately 402 MFLOPs and an average inference latency of 5.01 ms.

In a low SNR environment scenario, as shown in Table \ref{scn2}, the PCCs for all classification models are around 50 \%. Among them, our proposed model exhibits the highest classification performance compared to other models. At lower SNR levels, the NAEL framework activates the ARN more frequently to compensate for increased noise level, which leads to an overall rise in FLOPs. In this scenario, ARN activation significantly increased, resulting in an average computational load of about 606 MFLOPs and an average inference time of 8.44 ms. The computational cost of our proposed model was also the lowest among those achieving high recognition accuracy within the comparison group. 

Furthermore, our proposed method outperformed the Vision Transformer in recognition performance despite having lower computational complexity. This is because the proposed model effectively captures local feature relationships, whereas the Vision Transformer relies on a global self-attention mechanism that may not be optimal for our dataset. In modulation recognition tasks, the key features in time-frequency images are often spatially correlated, meaning that important information is locally connected rather than widely dispersed. Conventional CNNs, such as those used in our model, excel at learning these local patterns hierarchically, making them more effective in such cases. Conversely, Vision Transformers divide the input into patches and learn global dependencies, which is beneficial for large datasets with widely scattered features but less effective when key information is densely connected\cite{VITdis}. Considering the size of our dataset, the self-attention mechanism in Vision Transformers struggles to generalize effectively, resulting in suboptimal recognition performance compared to our approach. 

In the third scenario, with extremely low signal power as shown in Table \ref{scn3}, the PCCs of the classification models are around 35 \%. Here, the proposed model shows superior performance with a PCC approximately 2.4 \% higher than the best-performing comparison model. This is because the NAN identifies significant noise-affected data in the input, utilizing both PRN and ARN adaptively for the final decision. However, in extremely low SNR environments, the activation frequency of ARN increases as noise impact on the input data becomes more significant. In this extremely low SNR scenario, the ARN activation rate was highest, triggered approximately 694 times (57.8 \%) out of 1,200 predictions, increasing the average computational load to about 660 MFLOPs and the average inference time to 9.35 ms. While the model maintains a balance between accuracy and efficiency, this increased invocation may impact latency-sensitive applications or energy-constrained systems. Nevertheless, due to its adaptive structure, the proposed model still maintains lower overall computational costs compared to conventional models, as ARN is only activated when necessary. Efficiency is achieved by effectively using only the PRN for classification in cases where data are likely unaffected by noise, thereby avoiding unnecessary activation of ARN.

\begin{figure}[]
\centering
\includegraphics[width= 0.48 \textwidth]{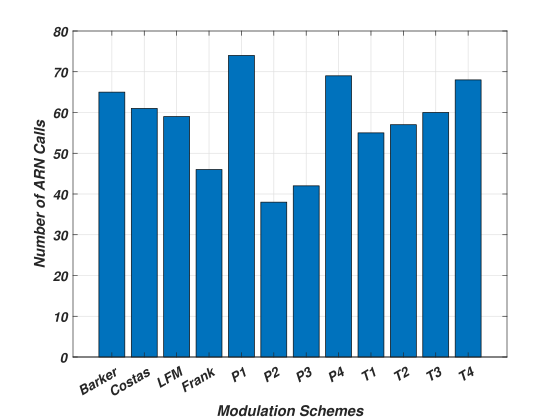}
\caption{ARN call count distribution across different modulation schemes in the third scenario.}
\label{fig_R1}
\end{figure}

Additionally, we examined whether the noise awareness capability of NAEL is affected by the diversity of data. Fig. \ref{fig_R1} illustrates the ARN call count distribution across different modulation schemes in the third scenario. The results show that while the call count varies, it does not exhibit a significant bias toward any specific modulation scheme. This suggests that while some variation exists, NAEL's noise awareness capability remains relatively stable across different modulation schemes. Moreover, since our dataset encompasses various noise and interference environments, as presented in Fig. \ref{fig_experiment}, the proposed method demonstrates robustness across diverse signal conditions, supporting its applicability in real-world scenarios.

In summary, our recognition model demonstrates a lightweight structure in high SNR environments where precise classification may be unnecessary. Moreover, our model achieves excellent modulation recognition performance in low SNR environments with minimal increase in computational cost. These findings underscore the adaptability of the proposed NAEL framework to varying conditions, enabling efficient modulation recognition in real-world scenarios.

\section{Conclusion}
This study proposes NAEL, a deep-learning framework for radar modulation recognition that adaptively adjusts its model structure based on noise conditions. The NAEL framework utilizes efficient modulation classification networks with high feature extraction capabilities and low computational costs. A key component, the noise-aware network within NAEL, evaluates noise impact on recognition, allowing adaptive selection of neural network structures for varying noise levels. Experimental analysis demonstrates that the proposed model outperforms existing classification models in recognition accuracy across diverse noise environments. Moreover, the proposed model exhibits lower computational demands compared to alternatives, a distinction magnified under high signal powers. These results position the NAEL framework as promising for ES systems requiring efficient resource utilization, rapid inference speeds, high accuracy, reliability, and interpretability.

\end{document}